\documentclass[12pt,preprint,usegraphicx]{aastex}

\def\aap{A\&A}

\newcommand\msun{\ifmmode{\hbox{M$_\odot$}}\else{M$_\odot$}\fi}
\newcommand{\reduceme}{\mbox{R\raisebox{-0.35ex}{E}D%
\hspace{-0.05em}\raisebox{0.85ex}{uc}\hspace{-0.90em}%
\raisebox{-.35ex}{{m}}\hspace{0.05em}E}}

\shorttitle{C,N abundances in cluster ellipticals}
\shortauthors{S\'{a}nchez--Bl\'{a}zquez et al.}

\begin{document}

\title{Differences in carbon and nitrogen abundances between field and cluster early-type 
galaxies}

\author{S\'{a}nchez--Bl\'{a}zquez P., Gorgas J., Cardiel N.\altaffilmark{1} and Cenarro J.}
\affil{Departamento de Astrof\'{\i}sica, Facultad de F\'{\i}sicas, Universidad Complutense de Madrid, 28040 Madrid, Spain}
\email{pat@astrax.fis.ucm.es}
\author{Gonz\'{a}lez J.J.}
\affil{Inst. de Astronom\'{\i}a, Universidad Nacional Aut\'onoma de M\'{e}xico, Apdo-Postal 70-264, M\'exico D.F, M\'exico}
\altaffiltext{1}{Also at Calar Alto Observatory, CAHA, Apdo. 511, 04044, Almer\'{\i}a, Spain}

\clearpage

\begin{abstract}
 Central line-strength indices were measured in the blue spectral region
for a sample of 98 early-type galaxies in different environments.
For most indices (Mgb and $\langle$Fe$\rangle$ in particular) ellipticals in rich clusters
and in low-density regions follow the same index-sigma relations.
However, striking spectral differences between field ellipticals and
their counterparts in the central region of the Coma cluster are found for
the first time, with galaxies in the denser environment showing
significantly lower C4668 and CN$_{2}$ absorption strengths. The most
convincing interpretation of these results is a difference in abundance
ratios, arising from a distinct star formation and chemical enrichment
histories of galaxies in different environments. An scenario in which
elliptical galaxies in clusters are fully assembled at earlier stages
than their low-density counterparts is discussed.

\end{abstract}

\keywords{galaxies: abundances --- galaxies: evolution --- galaxies: clusters --- galaxies: formation --- 
galaxies: stellar content}

\section{Introduction}

Despite the observational and theoretical efforts of the last decades, the
evolutionary status of early-type galaxies is still an unsolved problem. The
stellar populations of nearby ellipticals preserve a record of their formation
and evolution. In particular, the study of their element abundance ratios
should be a powerful discriminant between different star formation histories
(e.g. \citealt{W98}). However, this last approach is still at its infancy.  The
pioneering works of late 1970s already revealed that abundance ratios in
early-type galaxies are often non-solar (\citealt{Ocon76}; \citealt{Pet76}).
Since then, several studies have provided compelling evidence of Mg/Fe
overabundances in massive ellipticals as compared with the solar ratio
(\citealt{b42}; \citealt{b41}; \citealt{b60}), which have been interpreted in
the light of several possible scenarios based on the understanding that Mg is mainly produced in
type II supernovae (\citealt{Fab92}; \citealt{b42}; \citealt{Mat94}). However, and in contrast with the above findings, another
$alpha$ element, namely Ca, seems to be underabundant in ellipticals
(\citealt{Ocon76}; \citealt{Sag02}; \citealt{Cen03}; \citealt{Tho03}), challenging
current chemical evolution models (\citealt{Mat94}; \citealt{Moll00}). 

Several authors (\citealt{W98};\citealt{b37}) have also noted a strengthening
in other absorption line-strengths, in particular in the IDS/Lick C4668 and
CN$_{2}$ indices, when compared with stellar population models
predictions. The variations of these two indices are mainly driven by C and N
(in the case of CN) abundances (\citealt{Tri95}), which could suggest a
possible enhancement of these two elements relative to Fe when compared with
the solar values. In contrast with Mg, C and N are mainly produced in low- and
intermediate-mass stars (\citealt{Ren81}; \citealt{Chia03}, although there are
recent suggestions that most of the C should come from massive stars). During
the AGB phase, these stars eject into the ISM significant amounts of $^{4}$He,
$^{12}$C, $^{13}$C and $^{14}$N, enriching the medium from which new stars
will be formed. Therefore, it seems difficult to simultaneously reproduce  the
abundances of all these elements with a simple chemical evolution scenario.

The problem of C and N abundances has been more thoroughly studied in the
field of globular clusters. An interesting puzzling problem is the
existence of a CN dichotomy in Galactic and M31 globular clusters
(\citealt{b15}). Although this is a controversial issue, recent works
(\citealt{Har03}) tend to favor the scenarios of different abundances in the
parental clouds against the ones that predict abundance changes produced
internally by the evolution of the stars (see \citealt{Can98} for a review).

Given the expected sensitivities of relative abundances to the star formation
history of ellipticals, their study in galaxies within different environments
should help to discriminate between different formation and evolution models.
For instance, hierarchical scenarios predict that ellipticals in rich clusters
assembled completely at high redshift ({\it z}$>$3), whereas field ellipticals
may have experienced an elapsed and more complex star formation history
(\citealt{Kauff98}). However,  very little is known about the dependence of the
relative abundances on environment. One piece of information is that there is
no difference in the [Mg/Fe] ratio between cluster and field elliptical
galaxies (\citealt{Jor99}; \citealt{Kun02}).  In this letter, we study the
behaviour of several Lick/IDS indices (see \citealt{Worea94} for definition) in a sample of low and
high density environment galaxies (LDEG and HDEG respectively) and,
surprisingly, we do find systematic differences in the strength of C and CN
features.

\section{Observations and data analysis}

\subsection{Observations}

Long--slit spectra of 98 early--type galaxies in different environments were
taken in four observing runs with two different telescopes.  The sample
comprises 59 galaxies from the field and the Virgo cluster (LDEG), and 34 galaxies
from the central region of the Coma cluster (HDEG), spanning a wide range of
absolute magnitudes ($-22.5<M_{\rm B}<-16.5$, using $H_0 =
75$~km~s$^{-1}$~Mpc$^{-1}$), and central velocity dispersions $40< \sigma <
400$~km~s$^{-1}$ (from dwarf ellipticals, in Virgo and Coma, to giant galaxies).

In the first two runs (1998 January and 1999 August) we used the 3.5m
telescope at Calar Alto Observatory (Almer\'{\i}a, Spain), employing the Twin
Spectrograph. The observations of the third and fourth runs (1999 March and
2001 April) were carried out with the 4.2m WHT at the Roque de los Muchachos
Observatory (La Palma, Spain) using the ISIS spectrograph. Spectral
resolutions range from 2.6~\AA\ and 4.0~\AA\ (FWHM) for LDEG to 8.6~\AA\
for HDEG, in a spectral range $\lambda\lambda 3600$--5400~\AA. 
Exposures times of 1200--3600 secs per galaxy allowed us to obtain central
spectra with signal-to-noise ($S/N$) ratios (per \AA) ranging from 25 to 250. We
also observed several galaxies in common between runs to ensure that the
measurements were in the same system. $85$ stars from the IDS/Lick
library were included to transform the measured line-strength indices to the Lick system.

Standard data reduction procedures (flat-fielding, cosmic ray removal,
wavelength calibration, sky subtraction and fluxing) were performed with
{\reduceme} (\citealt{Car99}), which allowed a parallel treatment of data and
error frames and provided an associated error file for each individual data
spectrum.

\subsection{Velocity dispersion and line-strength indices}

For each galaxy, central spectra were extracted by coadding within a
standard metric aperture size corresponding to 4 arcsec projected at the
distance of the Coma cluster (simulating in this way a fixed linear 
aperture of length 1.8 kpc in all the galaxies). Velocity dispersions were determined 
using the MOVEL and OPTEMA
algorithms described in \citet{G93}.  We measured the Lick/IDS line-strengths
indices, although only CN$_{2}$, Mgb, C4668 and $<$Fe$>$ are presented here
(the rest of the indices, with a more detailed explanation of data handling,
will be presented in a forthcoming paper). All the indices were transformed to
the Lick spectrophotometric system using the observed stars and following the
prescriptions in J. Gorgas, P. Jablonka, \& P. Goudfrooij (2003, in preparation); see
also \citet{b33}.  Using galaxies in common with \citet{b34} and with repeated
observations between runs, we double-checked that there are no systematic
errors in the indices of galaxies observed in different runs.  All the indices
presented in this work are transformed into magnitudes, following
\citet{KunPhD}: $I'({\rm mag})=-2.5\log(1-I({\rm \AA})/\Delta\lambda)$, where
$\Delta\lambda$ is the width of the index bandpass.

\subsection{Results}

\begin{figure}
\rotatebox{-90}{
\plotone{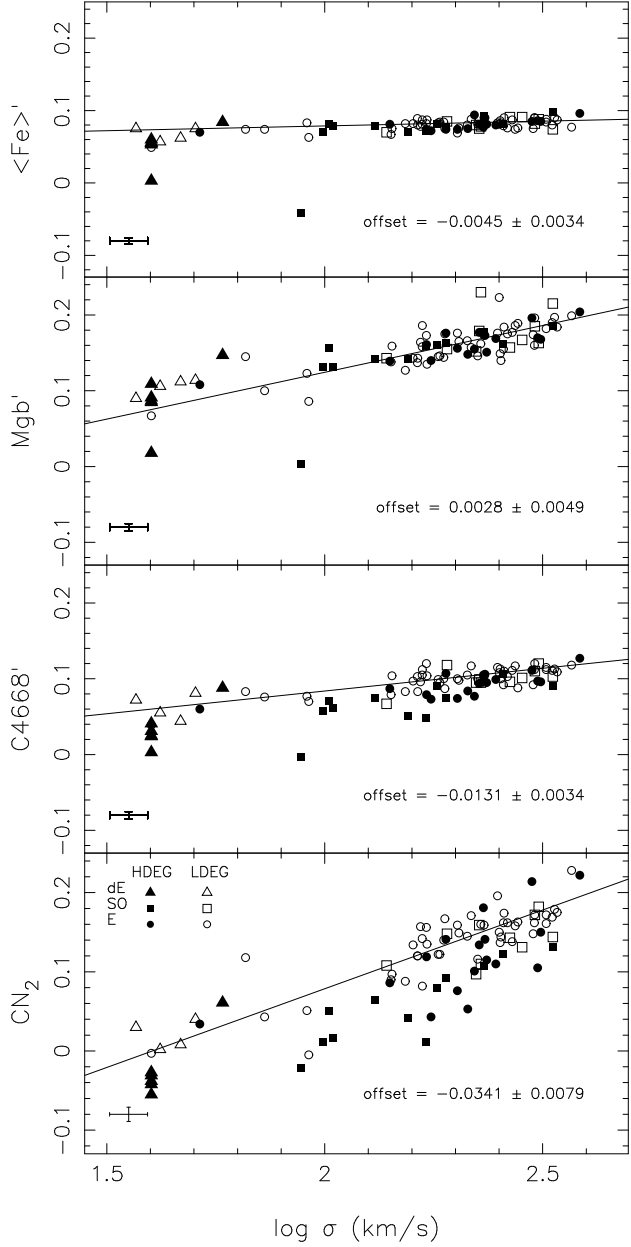}}
 \figcaption{Relations of the line-strength indices analyzed in this paper with
 central velocity dispersion for HDEG and LDEG. LDEG are represented with open
 symbols whereas filled symbols correspond to galaxies from the Coma
 cluster. Dwarfs ellipticals are plotted with triangles, S0 with squares and E
 galaxies with circles. Lines represent error-weighted least-squares linear
 fits to the LDEG subsample.  Typical errors in the indices and $\sigma$'s are
 included at the bottom-left of each panel.  Labels indicate the mean offsets,
 and their corresponding errors, of the HDEG with respect to the fits.
 \label{todas.sigma.letter}}
\end{figure}

In Fig.~\ref{todas.sigma.letter} we present the  CN$_{2}$, C4668$'$,
Mgb$'$, and $\langle$Fe$\rangle'$ indices versus velocity dispersion ($\sigma$) for
the 98 galaxies of the sample.  The spectra of three dwarf galaxies from the
Coma sample did not have enough $S/N$ to derive a reliable
measurement of $\sigma$ and a typical value of 40~km~s$^{-1}$ was assumed.
Also, for the rest of the galaxies with $\sigma<60$~km~s$^{-1}$, we adopted
$\sigma$ values from \citet{b50} and \citet{Guz03}.
It is apparent from this figure that galaxies located in low
and high density environments show systematic differences in C4668$'$ and
CN$_{2}$, being the indices in HDEG systematically lower than in LDEGs. On the
other hand, both galaxy subsamples seem to follow similar relationships in the
$\langle$Fe$\rangle'$ and Mgb$'$ versus $\sigma$ diagrams.

To quantify the possible systematic differences, we have performed a linear
least-square fit to the LDEG subsample, and have measured the mean offsets
(weighting with errors) of the Coma galaxies (HDEG) from the fits. These
differences and their errors are indicated within each panel, confirming the
high significance of the systematic offsets in CN$_{2}$ and C4668$'$.
  
\begin{figure}
\rotatebox{-90}{
\epsscale{0.7}
\plotone{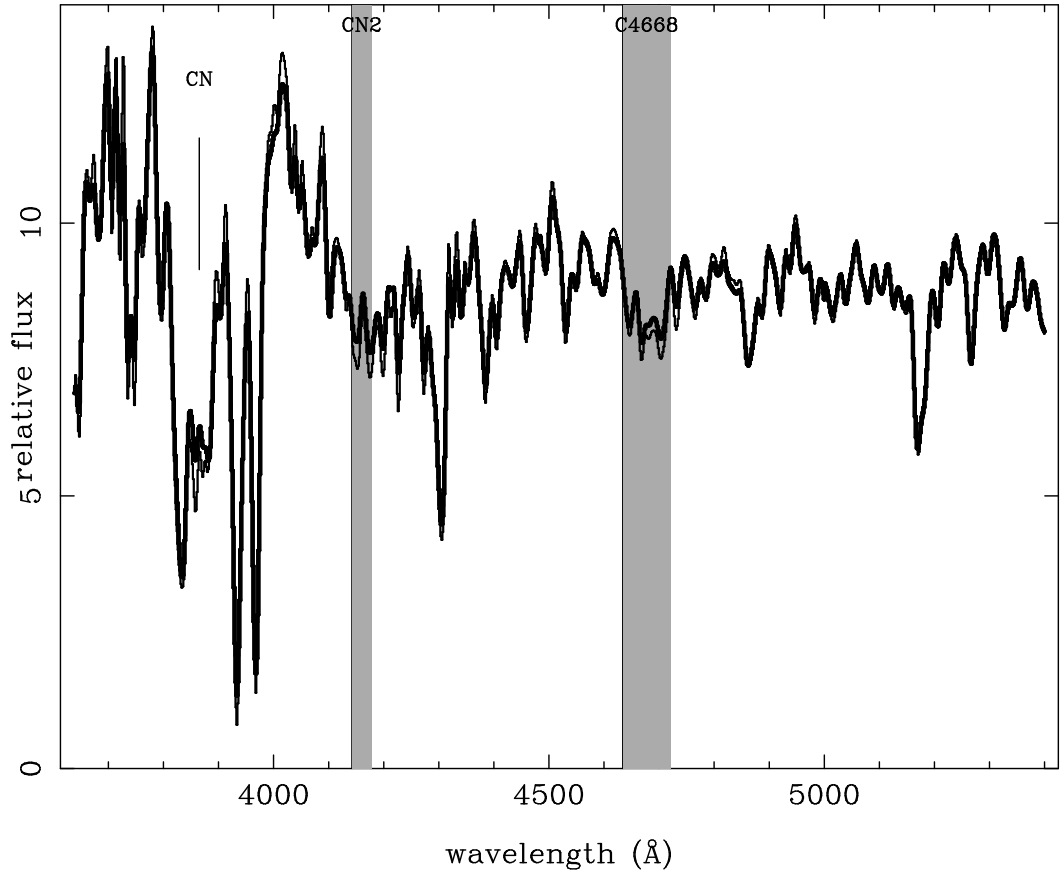}}
\figcaption{Coadded spectra for LDEG and HDEG, represented with thin and thick
 lines respectively, in our spectral range. See text for details. The position
 of the CN band around $\lambda4177$~\AA, and the central bandpasses of the
 CN$_{2}$ and C4668$'$ indices are indicated. Besides these bands, a very consistent
difference in the CN violet system (3850-3890 \AA) is also evident.
\label{espectro-figure}}
\end{figure}

These systematic differences are also visible directly in the spectra of the
galaxies. For illustration, Fig.~\ref{espectro-figure} shows the coadded
spectra of LDEG (thin line) and HDEG (thick line) in the range
$150<\sigma<250$~km~s$^{-1}$, previously shifted to the same radial velocity
and broadened to the maximum $\sigma$. The $S/N$ ratios of the two combined
spectra are above 300.  It is evident that the offsets found in
Fig.~\ref{todas.sigma.letter} are due to real enhancements of the CN band at
$\lambda4177$~\AA\ and the C$_{2}$ Swan band at 4735\AA\ in LDEG compared to
HDEG.  Note that this effect is also quite remarkable in the CN band around
$\lambda3865$~\AA\ (not included in the Lick system).

\section{Discussion}

Prior to interpret the systematic differences in CN$_{2}$ and C4668$'$ as
variations on elements abundances, and since the indices are also sensitive to
other physical parameters, we have explored further possibilities:

(1) Given the gravity dependences of the CN$_{2}$ and C4668$'$ indices
(both are stronger in giant than in dwarf stars;
\citealt{Worea94}), a decrease in the giant/dwarf ratio in HDEG (with respect
to LDEG) would lead to lower index values. Using the models by Vazdekis et al. (1996), 
we have checked that a change in the exponent of the IMF from
0.80 to 2.80 would decrease CN$_{2}$ and C4668$'$ by 0.033 and 0.009 mag
respectively, while the expected changes in Mgb$'$ and $<$Fe$>$ would be of
0.005 and 0.003 in the opposite sense (assuming an age of 10~Gyr and solar
metallicity). These predictions are marginally consistent with our results
(note that the above offsets for Mgb$'$ and $<$Fe$>$ are compatible with the
measurements within the uncertainties),
thus we cannot reject this possibility. However, this result is in
contradiction with other studies (\citealt{b36}) which suggest a decrease in the
giant/dwarf ratio in LDEG compared to HDEG, and would imply important changes
in other observables. Measurements of the Ca triplet in the near-infrared
(with a high sensitivity to the IMF; see \citealt{Cen03}) should help to
discard or confirm this scenario.
 
(2) A difference in (luminosity-weighted) mean age between HDEG and LDEG could
also introduce systematic offsets in CN$_{2}$ and C4668$'$ between both galaxy
samples. Under this scenario, and using the models by V96, the observed
offsets could be accounted for if HDEG were
about 8 Gyr younger than LDEG. This possibility can be rejected since it would imply 
a decrease in Mgb$'$ and $<$Fe$>$ of 0.026 and 0.014 in HDEG compared with
LDEG, which is not observed. Furthermore, it contradicts previous findings which
suggest that HDEG are, in any case, older than LDEG (\citealt{Kun02}).

\begin{figure}
\rotatebox{-90}{
\epsscale{0.7}
\plotone{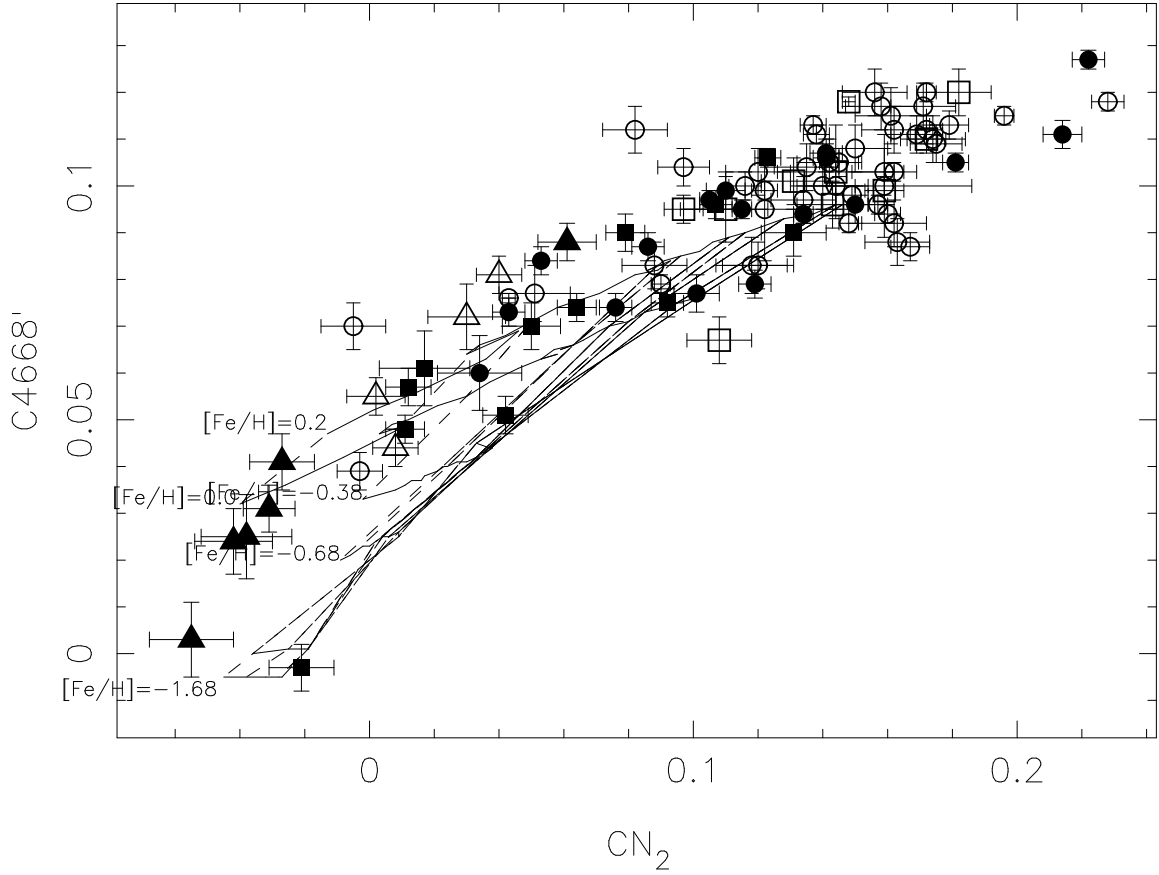}}
 \figcaption{C4668$'$ line-strengths versus CN$_{2}$ for the complete samples of
HDEG and LDEG. Overplotted are models by Vazdekis (1996).  Lines of constant
[Fe/H]= -1.68,-0.68,-0.38, 0.0, +0.2 are shown by solid lines. Dashed lines represent
models of constant ages, from 1 Gyr to 17.78 Gyr, increasing from 
left to right. Symbols are the same as in Fig.~\ref{todas.sigma.letter}.
 \label{index-index}}
\end{figure}

Thus, the most plausible explanation of our results is that variations in the
relative abundances of C and N with respect to Mg and Fe are responsible for
the observed offsets between galaxies in different
environments. Fig.~\ref{index-index} compares the observed CN$_{2}$ and
C4668$'$ with the predictions of stellar population models. This figure
clearly shows that the previously found overabundances of C and N only stand for LDEG, while
HDEG tend to exhibit relative abundances closer to the solar partition.

The C4668$'$ index is extremely sensitive to carbon changes, so little
variations in carbon abundance can change this index dramatically
(\citealt{Tri95}). However, the variations of CN are mostly controlled by N,
because extra C is readily incorporated into CO but extra N makes more CN
molecules. Therefore, a change in both C and N abundances is required to
explain our results. Besides, if extra carbon is easily incorporated into CO,
one should detect an enhancement in the strength of the CO bands in LDEG
compared to HDEG.  This effect has indeed been found by \citet{Mob96}
comparing the CO band at 2.3$\mu$m in galaxies from the field and from the
Pisces and Abell~2634 clusters.  They interpreted this difference as an
evidence of intermediate-age stellar population in LDEG (through a major
contribution of AGB stars). Although we do not discard a larger contribution
of younger populations in LDEG compared to HDEG, our results in the blue
spectral range imply that their observations can be solely explained by a
relative abundance difference. In any case, the observed offsets can not be
due to an age effect alone (see above).

The most plausible scenario to explain the differences in relative abundances
is the one in which LDEG and HDEG have experienced different star formation
histories. In particular, since the ISM is progressively enriched in C and N by
stars of low and intermediate mass stars, HDEG should have been fully assembled
before the massive release of these elements. The hierarchical clustering
paradigm currently predicts that galaxies in
clusters formed at different epochs than LDEGs. If the time elapsed between the
assembling of the former and the later is enough to permit the C and N
enrichment of the ISM of the pre-merging building blocks in LDEG, the stars
formed in these galaxies during the merging events will be C and N
enhanced. The constancy of the iron-peak elements could be understood if, in
both environments, the mergers take place before type-I SN can significantly
pollute the ISM of the pre-merging blocks.  Additionally, HDEG could have
experienced a truncated star formation and chemical enrichment history
compared to a more continuous time-extended history for their counterparts in
low density environments.  However, under this hypothesis, there should be an
increase of magnesium (produced by type-II SN) in LDEG which is not
detected. One way to understand the constancy of the Mgb$'$ index could be to
invoke the IMF-metallicity relationship suggested by \citet{Cen03}, in
which the succesive episodes of star formation in LDEG would take place with
lower giant-to-dwarf ratios.
  
To conclude, we have noted for the first time a systematic difference in the
element abundance ratios of galaxies situated in different environments. These
differences impose strong constraints to models of chemical evolution and
galaxy formation. It is clear that more work is still needed to completely
understand the causes of the differences. In particular, it would be very
interesting to compare the CN$_{2}$ and C4668' indices between the central
regions and the outskirts of the Coma cluster, where \citet{Mob00} found
differences in the strength of the near-IR CO molecule. Also, the detailed
study of other dense clusters is highly needed to confirm whether this effect
is particular of the Coma cluster.

\section*{Acknowledgments}
We are indebted to R. Guzm\'{a}n for providing us with a catalogue of dwarf
elliptical galaxies in the Coma cluster.  
 The WHT is operated on the island of La Palma by the Royal Greenwich Observatory at the Observatorio del Roque de los Muchachos of the
Instituto de Astrof\'{\i}sica de Canarias. The Calar Alto Observatory is operated jointly by the Max-Planck-Institute f\"{u}r Astronomie, 
Heidelberg, and the Spanish Comisi\'{o}n Nacional de Astronom\'{\i}a. This
work was supported by the Spanish research project No.AYA2000-977.

\clearpage

\end{document}